\newcommand{\Om}{\Omega}			

\newcommand{\OmGVMp}{{\Omega_\mathrm{gvm}}}
\newcommand{\OmGVS}{{\Omega_\mathrm{gvs}}}

\newcommand{\dtA}{{\tau_\mathrm{gvm}}}

\newcommand{\tauGVS}{{\tau_\mathrm{gvs}}}

\newcommand{\sinc}{{\rm sinc}}

\newcommand{\nn}{\nonumber}
\newcommand{\bsub}{\begin{subequations}}
\newcommand{\esub}{\end{subequations}}
\newcommand{\beq}{\begin{equation}}
\newcommand{\eeq}{\end{equation}}
\newcommand{\beqa}{\begin{eqnarray}}
\newcommand{\eeqa}{\end{eqnarray}}
\newcommand{\beql}{\begin{subequations}\begin{eqnarray}}
\newcommand{\eeql}{\end{eqnarray}\end{subequations}}
\documentclass[aps, amsmath, showpacs, longbibliography]{revtex4}
\usepackage[english]{babel}
\usepackage{amsmath} 
\usepackage{graphicx}
\usepackage{hyperref}
\usepackage{float}   
\usepackage{verbatim}  
\usepackage{braket}
\usepackage{color}
\usepackage{cases}
\usepackage{natbib}
\begin{document}
\title{ Squeezing and EPR correlation in the Mirrorless Optical Parametric Oscillator}
\author{ A.~Gatti$^{1,2}$, T.~Corti$^1$, E.~Brambilla$^{1}$ }
\affiliation{ $^1$ Istituto di Fotonica e Nanotecnologie del CNR, Piazza Leonardo da Vinci 32, Milano, Italy, 
$^2$ Dipartimento di Scienza e Alta Tecnologia dell'Universit\`a dell'Insubria, Via Valleggio 11 Como, Italy.}
\begin{abstract}
This work analyses the quantum properties of counter-propagating twin beams generated by a Mirrorless Optical Parametric Oscillator in the continuous variable regime.  Despite the lack of the filtering effect of a cavity, we show that  in the vicinity of its threshold  it may generate high levels of  narrowband squeezing and Einstein-Podolsky-Rosen (EPR)  correlation, completely comparable to what can be obtained in standard optical parametric oscillators.
\end{abstract}
\maketitle
\begin{figure}[ht]
     \includegraphics[width=0.55\textwidth]{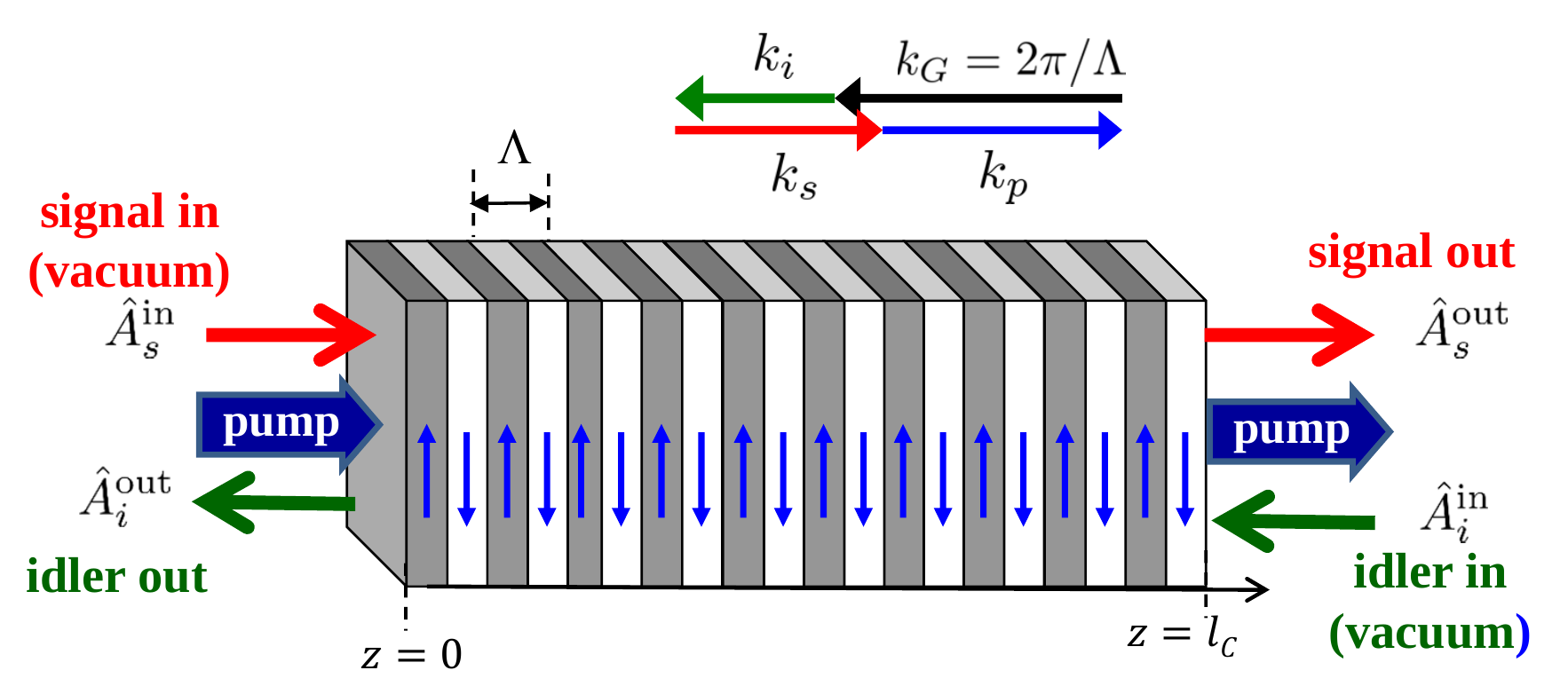}
           \caption{(Color online)Scheme of backward PDC, taking place in a $\chi^{(2)}$  crystal periodically poled with a submicrometer period $\Lambda \approx \lambda_p/n_p$. Quasi-phase matching  [Eq.\eqref{eq:pm}] requires then that the idler field is generated in the backward direction with respect to the signal and pump.}
\label{fig_scheme}
\end{figure}

%
Backward parametric down-conversion (PDC),    
where one of the twin beams back-propagates with respect  to the pump laser source (Fig.\ref{fig_scheme}), 
 is gaining an increasing  attention in the quantum optics community. In the spontaneous  regime it  has a natural potentiality to 
 generate high-purity and narrowband  heralded single photons \cite{Christ2009,Gatti2015,Brambilla2017}, a highly desirable and non trivial goal, which in  the standard  co-propagating geometry can be realized only at  specific tuning points.
A second appealing feature  is the presence of a threshold   pump intensity, beyond which the system makes a transition to coherent oscillations, i.e. it behaves as a {\em Mirrorless Optical parametric Oscillator} (MOPO)\cite{Canalias2007}. Responsible of this critical behaviour is the feedback mechanism established by back-propagation and stimulated down-conversion. Ref.  \cite{Corti2016} analysed the critical behavior of  twin beams below threshold, enlightening the role of the  quantum correlation of photon pairs in creating  the feedback necessary to the onset of a classical coherence above threshold. 
\par
In this work we turn our attention to the quantum properties of the source in the continuous-variable regime, so far unexplored, namely its potentiality to generate  EPR-correlated    beams in the vicinity of the threshold. 
EPR correlation \cite{Einstein1935,Reid1989, Ou1992}, i.e. nonclassical correlations in a pair of non-commuting field quadratures, and their associated squeezing, are features of the two -mode squeezed state produced  by any down-conversion process (see e.g.\cite{Knight2005a}).
 However, squeezed light generated in the standard single-pass configuration is  in general multimode, which is often undesirable for applications \cite{Lemieux2016} Moreover  high levels of squeezing are hard to be generated and detected  (see e.g. \cite{LaPorta1991}, but also  \cite{Eto2008,Kaiser2016} for recent 
achievements in this sense).  The typical solution is to recycle the parametric light  in an optical resonator, which at the same time enforces the nonlinearity and  produces a sharp modal filtering, i.e. to build an optical parametric oscillator (OPO).  Remarkably, this work will show that counterpropagating twin-beams, despite the lack of the filtering effect of the cavity,  exhibit high levels of narrowband  EPR correlation, completely comparable to what can be obtained in standard subthreshold  OPOs.
The role of the cavity is in the MOPO played by the distributed feedback mechanism\cite{Corti2016}, which creates a threshold where, similarly to the OPO, the quantum noise in principle diverges in some observables, allowing then noise suppression in their conjugate observables. 
Once technical challenges involved in its realization are overcome, this source may then represent a robust and compact  alternative to the  OPO.
\par 
The backward geometry  requires  a  sub-micrometer poling   of the  $\chi^{(2)}$ materials, which explains why after the first theoretical prediction \cite{Harris1966},  this source had to wait  forty years before being realized  \cite{Canalias2007}. 
We consider the scheme in Fig.\ref{fig_scheme},  in which the laser pump  at frequency $\omega_p$ and the signal  at frequency $\omega_s$ co-propagate along the $+z$ axis in the  nonlinear medium, while the idler at frequency $\omega_i=\omega_p-\omega_s$ 
back-propagates in the $-z$ direction.  Quasi-phase matching (i.e. the generalized momentum conservation)  is realized when
their corresponding  wave numbers $k_j=\omega_j \, n_j(\omega_j)/{c}$  satisfy 
\beq
k_s -k_i = k_p - m \frac{2\pi}{\Lambda}  \qquad m=1,3 \hdots 
\label{eq:pm}
\eeq
where $\Lambda$  is the poling period, and $ n_j $  the refraction indexes. 
First order interactions then require  $\Lambda \simeq \lambda_p/n_p $.
\par
Our  quantum model for this configuration was described in Refs.\cite{Corti2016,Gatti2015} 
(see also \cite{Suhara2010}).
 As in the former literature,   we restrict to a purely temporal description, assuming either
 waveguiding or a small collection angle.
Below the MOPO threshold the depletion of the pump laser is negligible, and it  can be described as  a   classical field  of constant  amplitude  along the sample.  Assuming in addition  that the pump is CW,   it is simply described  by its complex amplitude $\alpha_p=|\alpha_p|e^{i\phi_p}$.  
The strength of the parametric interaction is then characterized by the dimensionless gain  
\beq
g= \sqrt{2\pi} \chi |\alpha_p| l_c
\eeq
where $\chi$ is proportional to the $\chi^{(2)}$  susceptibility of the medium and $l_c$ is the crystal length.  In terms of this parameter the MOPO threshold occurs \cite{Ding1996} at 
\beq
g=g_{\mathrm{thr}} = \frac{\pi}{2}
\eeq

 The signal and idler waves are instead described by quantum field operators $\hat{A}_s(\Omega,z)$ and  $\hat{A}_i(\Omega,z)$, 
for two wavepackets centered around 
the respective reference frequencies $\omega_s$ and $\omega_i$ satisfying  quasi-phasematching (\ref{eq:pm}) (capital $\Omega$ is the  offset from the $\omega_j$).
As detailed in \cite{Corti2016},  the model is formulated  in terms of linear propagation equations coupling only frequency conjugate modes
$\omega_s +\Om$, $\omega_i -\Om$   of  the twin beams, whose solution gives a  transformation linking the output operators   $\hat{A}_s^{\text{out}} =  \hat{A}_s(z=l_c )  $, $\hat{A}_i^{\text{out}}= \hat{A}_i( z=0 )$
 to the input ones  (Fig.\ref{fig_scheme}), assumed in the vacuum state. Notice that in this geometry the boundary conditions are not the standard ones,  because  the signal and idler  fields exit  from the opposite end faces of the slab.  
The input-ouput relations are then the Bogoliubov transformation, characteristic of processes where particles are generated in pairs:
\bsub
\begin{align}
\hat{A}_s^{\text{out}}(\Omega)&=U_s(\Omega)\hat{A}_s^{\text{in}}(\Omega)+V_s(\Omega)\hat{A}_i^{\text{in} \dagger}(-\Omega)\\
\hat{A}_i^{\text{out}}(-\Omega)&=U_i(-\Omega)\hat{A}_i^{\text{in}}(-\Omega)+V_i(-\Omega)\hat{A}_s^{\text{in}\dagger}(\Omega).
\end{align}
\label{inout}
\esub
The coefficients $U_j(\Omega)$ and $V_j (\Omega)$ are  the trigonometric functions  \cite{Corti2016}:
\bsub
\begin{align}
U_s(\Omega)&=e^{i k_s l_c}e^{i\beta(\Omega)}\phi(\Omega)\\
V_s(\Omega)&=e^{i (k_s-k_i) l_c}g e^{i\phi_p}\frac{\sin\gamma(\Omega)}{\gamma(\Omega)}\phi(\Omega)\\
U_i(-\Omega)&=e^{i k_i l_c}e^{i\beta(\Omega)}\phi^*(\Omega)\\
V_i(-\Omega)&= ge^{i\phi_p}\frac{\sin\gamma(\Omega)}{\gamma(\Omega)}\phi^*(\Omega) \qquad \text{with} \\
\phi(\Omega)&=\frac{1} {\cos\gamma(\Omega)-i\frac{\bar{\mathcal{D}}(\Omega)l_c}{2\gamma(\Omega)}  \sin\gamma(\Omega) } \nn  \\
\gamma(\Omega)&=\sqrt{g^2+\frac{{\mathcal{D}}^2(\Omega)l_c^2}{4}},  \label{Gamma} 
\end{align}
\label{UV}
\esub
In these expressions: 
\begin{align}
{\mathcal{D}}(\Omega) &=k_{s}(\Omega)-k_{i}(-\Omega)-k_p+k_G 
\label{DD}
\end{align}
is the phase mismatch for two frequency conjugate signal-idler components, $k_j(\Omega)$ being the wavenumber of $j-$th wave  at frequency $\omega_j+\Omega$ ($j=s,i$). 
The phase 
\beq
\beta(\Omega)=[k_s(\Omega)+k_i(-\Omega)-(k_s+k_i)]\frac{l_c}{2}
\label{eq:phase}
\eeq
is  a global propagation phase. 
Notice that the coefficients $U_j(\Omega)$ and $V_j(\Omega)$ diverge when approaching  the MOPO threshold $g=\pi/2$, and, 
as can be easily checked, they  satisfy  the unitarity conditions: 
$
|U_j(\Omega)|^2-|V_j(\Omega)|^2=1$, and $U_s(\Omega)V_i(-\Omega)=U_i(-\Omega)V_s(\Omega)$
\par
\begin{figure}[h]
     \includegraphics[width=0.65\textwidth]{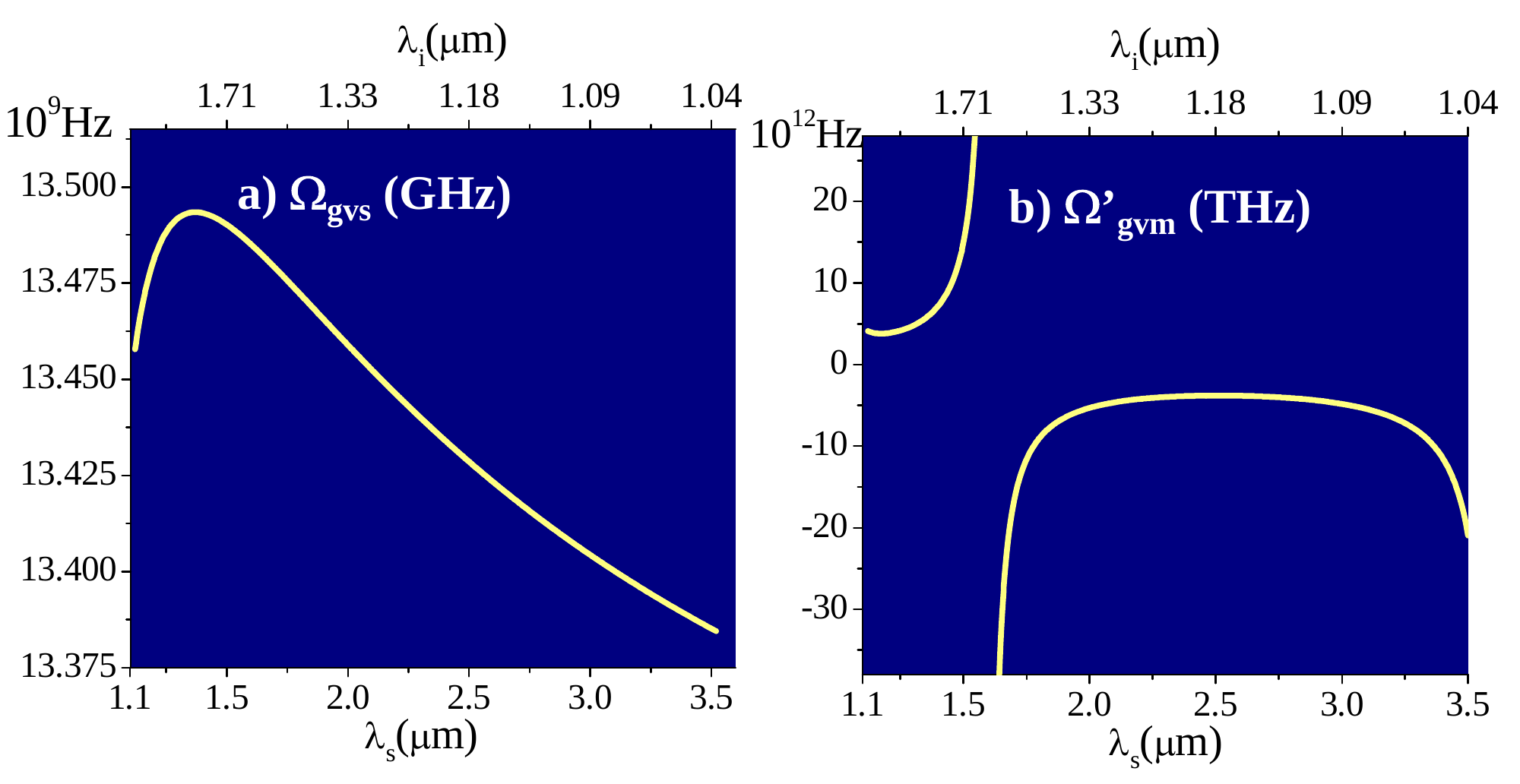}
           \caption{(Color online) Comparison between the two spectral scales of the MOPO for PPLN pumped at 800 nm, $l_c=1 cm$. a) $\OmGVS= \frac{2}{l_c} ({k'_s+k'_i})^{-1} \simeq 10 $Ghz is the narrow  MOPO bandwidth;  b) $\OmGVMp
= \frac{2}{l_c} ( {k'_s-k'_i})^{-1}  $ in the  range 5Thz  or more  is the broader GVM bandwidth. For each  $\lambda_s, \lambda_i$ the poling period is chosen to realize phase-matching  according to Eq.\eqref{eq:pm}.  }
\label{fig_comparison}
\end{figure}

Unlike the co-propagating case, this configuration  is characterized by narrow spectral bandwidths \cite{Canalias2007,Gatti2015,Corti2016}. Therefore,  it is legitimate to  retain only the  first order of the Taylor expansions of the wavenumbers $k_j(\Om)$, so that 
\begin{align}
\frac{ {\mathcal{D}}(\Omega)l_c}{2}
& \simeq \frac{l_c}{2}(k_s'+k_i')\Omega: = \frac{\Om} {\OmGVS}  
\label{linear}\\
\beta(\Omega)&\simeq(k'_s-k'_i)\frac{l_c}{2}\Omega=
\frac{\Om} {\OmGVMp}, 
\label{linear_beta}
\end{align}
where 
 $k'_j = \frac{dk_j }{d\Om}  |_{\Om=0}$ , and
\beq
  \Om_{\mathrm{gvs}}^{-1}  \equiv\tauGVS= \frac{1}{2}\left[\frac{l_c}{v_{gs}}+\frac{l_c}{v_{gi}}\right].
\eeq
is a {\em long}  time scale characteristic of counterpropagating interactions, on the order of the  transit time of light along  the slab, 
 involving the {\em sum} of the inverse group velocities $v_{gj }= 1/k'_j$  \cite{Corti2016, Gatti2015}.    In the spontaneous regime,   it  defines  the  correlation time  of twin photons, while its  inverse $\OmGVS$ gives the narrow width of their spectrum, which  becomes even narrower  in the stimulated regime and ideally shrinks to zero on approaching threshold \cite{Corti2016}.
Conversely 
\beq
 \Om_{\mathrm{gvm}}^{-1}  \equiv \dtA=\frac{l_c}{2 v_{gs}}-\frac{l_c}{2 v_{gi}}
\label{DtA}
\eeq
is a {\em short} time scale related to the group velocity mismatch (GVM), and produces a small temporal offset between the signal and idler wave-packets. Clearly,   $|\OmGVMp| \gg  \OmGVS $ for any tuning conditions (see Fig.\ref{fig_comparison} for  a comparison in the case of  LiNbO$_3$).  
Within these linear approximations the coefficients $U_j(\Om) $ and $V_j(\Om)$  basically depend on the frequency only through the ratio $\frac{\Omega^2}{\Om^2_{\mathrm gvs}}$, because 
$
\gamma(\Om) \simeq \sqrt{g^2 + \frac{\Omega^2}{\Om^2_{\mathrm gvs} }}, 
\label{gamma}
$
while
the phase  $\beta(\Om)$ in \eqref{linear_beta} varies  on the slow scale $ |\OmGVMp |\gg  \OmGVS  $ and  remains close to  zero  in the spectral region where $U_j(\Om) $ and $V_j(\Om)$ take non trivial values. 
\par
Several properties of  the state of the MOPO below threshold depend  solely on the Bogoliubov  form \eqref{inout} of the transformation,   so that they are common 
 to any  linear  process of photon-pair generation.
In particular,  
if one introduces the sum and difference between frequency conjugate components of the twin beams: 
$
\hat C_{\pm} (\Om) = \frac{1} {\sqrt 2} [ \hat A_s^{\mathrm out} (\Om) \pm  A_i^{\mathrm out} (-\Om) ], 
$
then  
the transformation \eqref{inout} decouples into two independent {\em squeeze} transformations\cite{Gatti2017}. 
The  $\pm$ modes  are thus  individually squeezed,  and their  squeezing ellipses turn out oriented along  orthogonal directions.  As well known, this implies the simultaneous presence  of correlation and anticorrelation in two orthogonal quadrature operators of the twin beams \cite{Reid1989, Ou1992}.
\par
\begin{figure}[ht]
     \includegraphics[width=0.75\textwidth]{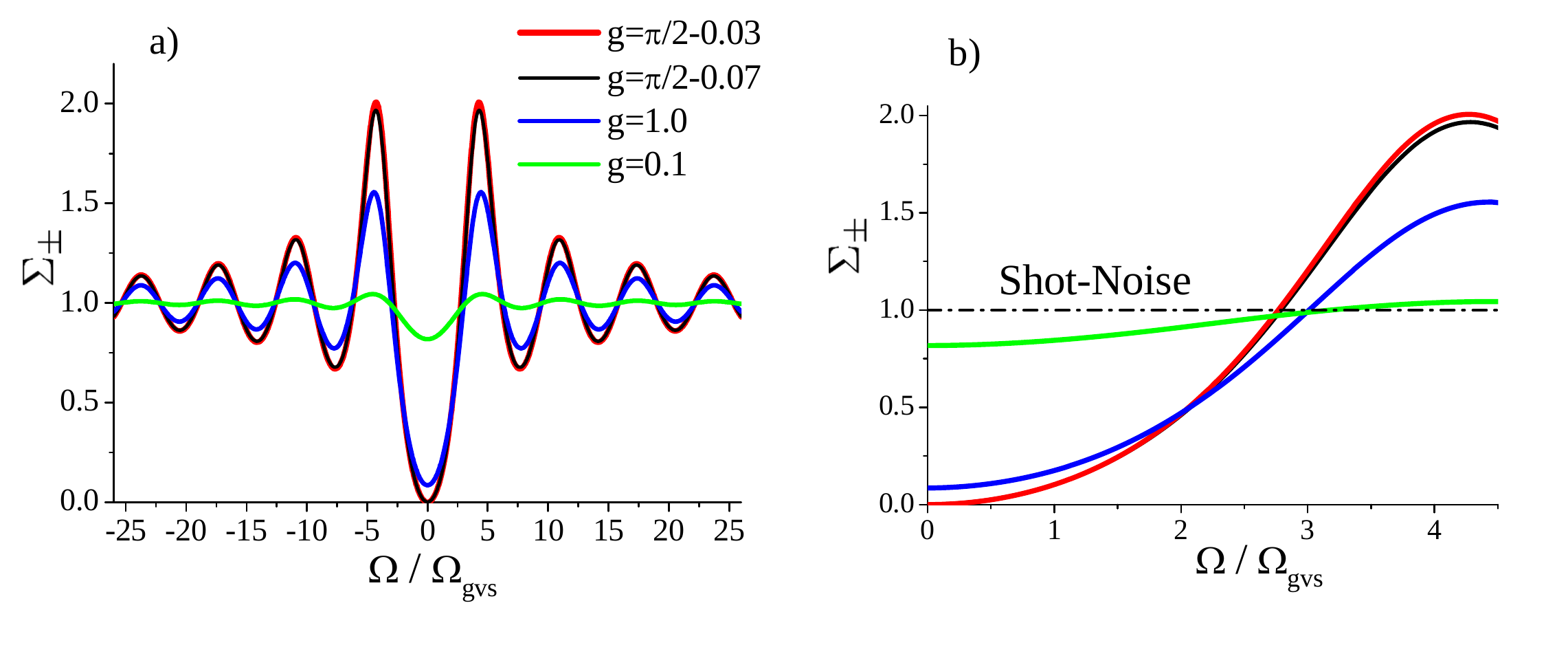}
           \caption{(Color online) Squeezing spectra $\Sigma_{\pm} (\Omega) $  \eqref{Sigmapm2}, and degree of EPR correlation between the MOPO twin beams, as a function of  $\Omega/\OmGVS$, for $\phi_s+ \phi_i$  fixed as in Eq.\eqref{phifix2}. $\Delta t = 0$.The inset  (b)  is a detail  of the minima. }
\label{fig_squeezing}
\end{figure}

 
In order to characterize the 
 amount of squeezing and EPR correlation generated in this specific configuration, let us consider    the quadrature operators   for the individual signal and idler  fields in the time domain
\begin{align}
\hat X_j (t) &= \hat A_j^ {\mathrm out}  (t)e^{-i \phi_j } + \hat A_j^ {\mathrm out \, \dagger }  (t)e^{i \phi_j},  
\label {Xtime}  \\
\hat Y_j (t) &= \frac{1}{i}  [\hat A_j^ {\mathrm out} (t)  e^{-i \phi_j} - \hat A_j^ {\mathrm out \, \dagger }  (t)e^{i \phi_j}] \quad j=s,i
\label{Ytime}
\end{align}
The two orthogonal quadratures do not commute   $[ \hat{X}_j(t), \hat Y_k (t')]= \delta_{j,k} \delta(t-t')$ and represent 
 incompatible observables.  
Notice that  their  Fourier transforms: 
$
\hat X_j (\Om) 
= \hat A_j^ {\mathrm out}  (\Om )e^{-i \phi_j } + \hat A_j^ {\mathrm out \, \dagger }  (-\Om )e^{i \phi_j} 
$ (which are not Hermitian and hence not observables)
 involve the  two symmetric spectral  components  $ \omega_j \pm \Om $ for each field.  We then introduce proper combinations of the signal and idler quadratures: 
\begin{align}
\hat X_{-} (t) &= \frac{ 1}{\sqrt 2} [\hat X_s(t) - \hat X_i (t - \Delta t)  ]\\
\hat Y_{+} (t) &= \frac{1}{\sqrt 2} [\hat Y_s(t) + \hat Y_i (t - \Delta t) ]
\label{Xpm_time}
\end{align} 
where   the delay  $\Delta t$ between the detection of the signal and idler arms can be used as an optimization parameter. 
Next, we characterize  the  noise in the sum or difference modes by the so-called {\em  squeezing spectra}
\begin{align}
\Sigma_\pm (\Omega) &= \int_{-\infty} ^{+ \infty}  d\tau \, e^{i \Om \tau}  \begin{cases}
\left\langle \delta \hat Y_+(t) \hat \delta Y_+ (t + \tau) \right\rangle  \\
\left\langle \delta \hat X_-(t) \hat \delta X_- (t + \tau) \right\rangle  
\end{cases}
\label{Sigmapm}
\end{align}
where   e.g. $\delta \hat X_- = \hat X_-  - \langle \hat X_-  \rangle = \hat X_-  $, because below the threshold the field expectation values are zero. 
These quantities describe the degree of  correlation ("-" sign) or anticorrelation ("+" sign) existing between the field quadrature operators of the twin beams at the two crystal output faces.  The value "1" represents the shot noise level, which corresponds to two uncorrelated  light beams. In the degenerate case $\omega_s=\omega_i$, one may also think of physically 
recombining  the two counterpropagating  beams on a beam-splitter,  in order to produce two independently squeezed  beams.\\
After some long but straightforward calculations, based on the input-output relations \eqref{inout},  one obtains
\begin{align}
\Sigma_{\pm} (\Om )  &= \frac{1}{2} \left \{ \left|  U_s(\Om)  - V_i^* (-\Om) e^{i \Om \Delta t} e^{i (\phi_s + \phi_i)} 
\right|^2  \right. \nn \\
&+ \left.  \left|  U_s (-\Om)  - V_i^* (\Om) e^{-i \Om \Delta t} e^{i (\phi_s + \phi_i)} 
\right|^2 \right\}
\label{Sigmapm2}
\end{align} 
Up to this point  we used only  the Bogoliubov form of the  relations \eqref{inout}, so that  Eq.\eqref{Sigmapm2} actually  holds for any PDC process. As expected for the EPR state, the degree of correlation and anticorrelation  in orthogonal quadratures are identical:  $\Sigma_{-} (\Om ) =
\Sigma_{+} (\Om )$.
The  two spectral terms at r.h.s of  Eq.\eqref{Sigmapm2}  
are present because detection of  the temporal 
  quadratures   \eqref{Xtime} probes the noise at  $\omega_j \pm \Om$ for each field. In the MOPO,  
these  terms can  be made identical by  setting $\Delta t= \dtA$, which exactly compensates the temporal offset of the twin beams. However, even in the absence of such optimization,  the two  terms   are  respectively  minimized by  choosing 
\begin{align} 
\phi_s + \phi_i    &= 2\theta (\pm \Om) 
 = \arg \left[  U_s(\pm \Om) V_i( \mp \Om) \right]  
\label{phiopt1} \\
&\simeq   k_s l_c + \phi_p   +\text{arg}\left[  \sinc \gamma (\Om) \right] \pm \frac { \Om }{\OmGVMp}
\label{phiopt2} \\
&\simeq   k_s l_c + \phi_p   +\text{arg}\left[  \sinc\gamma (\Om) \right]  
\label{phiopt}
\end{align}  
where the second line  uses the linear approximations \eqref{linear} and \eqref{linear_beta},  and the last line holds because 
$ \Om/\OmGVMp \approx 0$ within the spectral region of interest. 
With this choice
$
\Sigma_{\pm} (\Om ) \to \left[  \left| U_s(\Om) \right| -\left| V_i (-\Om)\right| \right]^2$
reaches its minimum value at any frequency, and the noise never goes above the shot noise level "1". 
The degree of EPR correlation/anticorrelation  $\Sigma_ \mp (\Om) $ 
is instead  plotted in Fig.\ref{fig_squeezing}
 for {\em fixed phase angles}, namely 
\beq
\phi_s + \phi_i  :=  2\theta (0)= k_s l_c + \phi_p 
\label{phifix2}
\eeq
 In this case,  the noise passes from below to above the shot noise at  $\Omega = \pm \OmGVS\sqrt{\pi^2 -g^2}  $,  where  $\sinc{\gamma(\Om)}$ changes sign. These values can be used to define a bandwidth of squeezing $  \Delta \Omega= 
\OmGVS\sqrt{\pi^2 -g^2} \approx 2.7  \OmGVS$ close to threshold. Some remarks are in order: 
i) The EPR correlation becomes asymptotically perfect as the MOPO threshold is approached, which can be realized only close to a critical point, because the noise in the quiet quadrature can be suppressed only at the expenses of a diverging level of noise in the orthogonal one. 
ii)  The squeezing remain significant at rather large distances from threshold, $\Sigma_\pm (0) \simeq 0.09 $ for $g=1$, which is $ 36\%$ below the MOPO threshold.
ii) Excellent levels of squeezing are present in the whole emission bandwidth,  that we remind is   smaller than $  \OmGVS$\cite{Corti2016}. This is
 in sharp contrast 
  with the  single-pass co-propagating geometry, where high squeezing is difficult to observe\cite{Eto2008}, and  the orientation  of the squeezing ellipses  varies rapidly inside the PDC bandwidth\cite{Gatti2017}. In contrast,  for the MOPO the  orientation of the  ellipses, defined by $\theta(\pm \Om)$,  remains practically constant  inside the  bandwidth  $ \OmGVS$ [see Eqs.\eqref{phiopt1}-\eqref{phiopt}]. This can be viewed as a consequence of the   {\em  long}   ($\tauGVS$)  and {\em short} ($\dtA$)  time scales  involved in the counterpropagating geometry. 
\par
\begin{figure}[t]
     \includegraphics[width=0.75\textwidth]{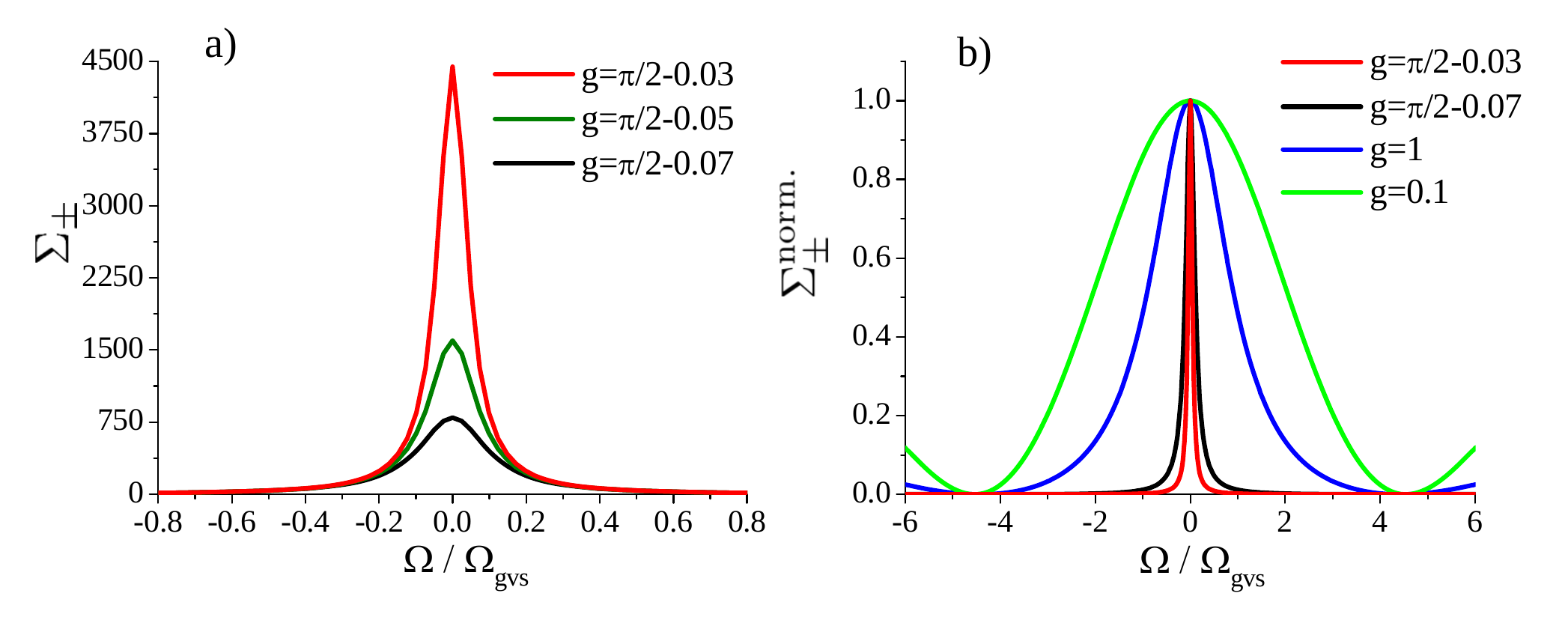}
           \caption{(Color online) Antisqueezing spectra $\Sigma_{\pm} (\Omega) $ \eqref{Sigmapm2}  as a function of $\Omega/\OmGVS$, for  phase-angles 
orthogonal to those in Fig.\ref{fig_squeezing}.  The curves in (b)  are normalized  inside $(0,1)$.}
\label{fig_antisqueezing}
\end{figure}

Fig.\ref{fig_antisqueezing} shows the antisqueezing of the sum or difference modes, which occurs for quadrature phases orthogonal to those in Fig.\ref{fig_squeezing}.  In this case the  noise diverges on approaching threshold, which is clearly reminiscent of the critical divergence of the MOPO spectra analysed in \cite{Corti2016}. The bandwidth of the antisqueezing spectra shrinks getting close to threshold (Fig.\ref{fig_antisqueezing}b), which again reflects the shrinking of the spectra and the critical slowing down of  temporal fluctuations close to the MOPO threshold\cite{Corti2016}.
\par
\begin{figure}[ht]
     \includegraphics[width=0.65\textwidth]{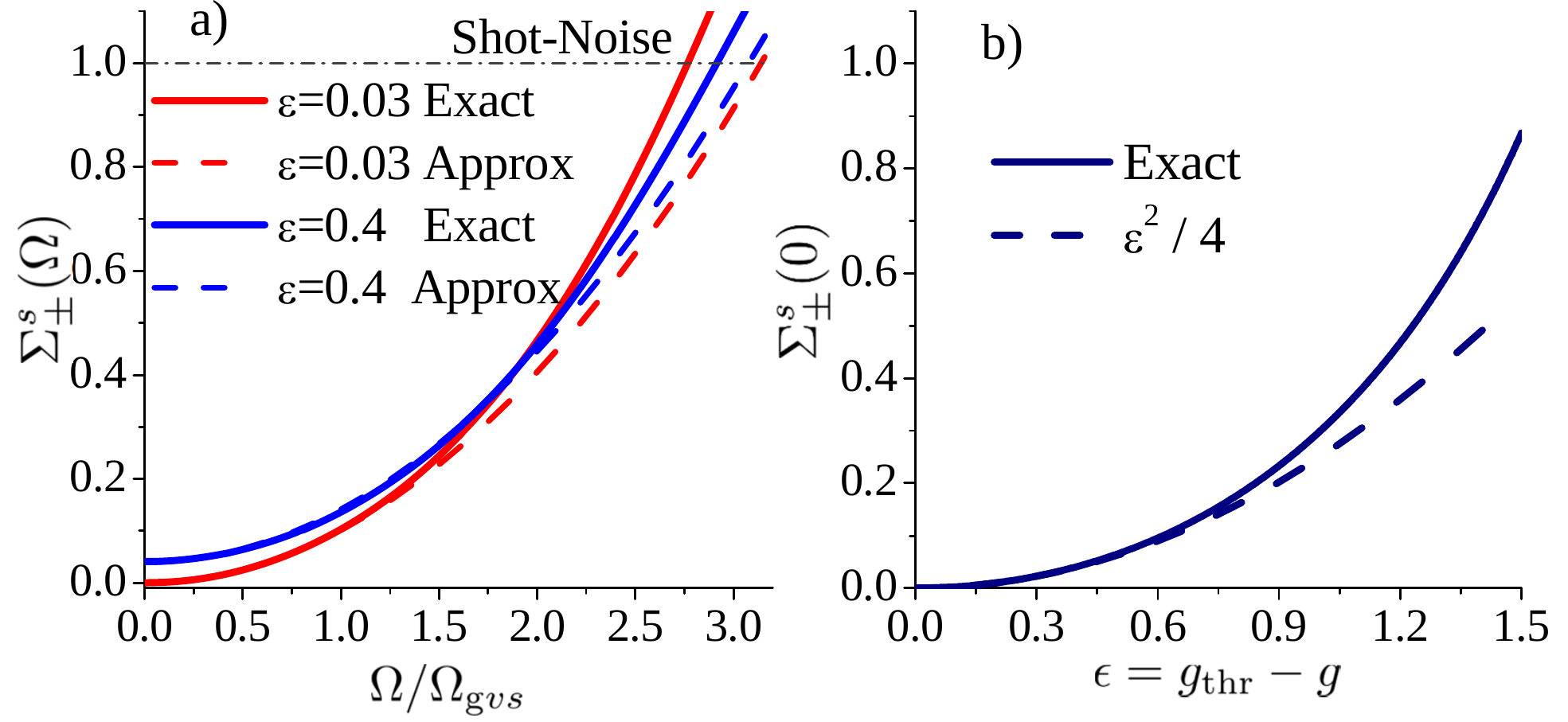}
           \caption{(Color online)Comparison between the exact results in Eq.\eqref{Sigmapm2},  and the approximated ones in Eq.\eqref{epsilon}. Squeezing  spectra (a) as function of frequency, and (b)  at zero frequency as a function of
the distance from threshold}.  
\label{fig_sigma0}
\end{figure}

The curves in  Figs.\ref{fig_squeezing}  and \ref{fig_antisqueezing} are in sense {\em universal} for the MOPO,  when plotted as a function 
of $\frac{\Omega}{\OmGVS}$, and to a very good approximation  
hold for any material and tuning conditions.     
This can be more clearly seen by  deriving explicit expressions of the noise spectra. By inserting the coefficients\eqref{UV}  in the general result\eqref{Sigmapm2}, 
using the linear approximation \eqref{linear} and neglecting the contribution of the slow phase $\beta(\Om)$, when $\phi_s+\phi_i$ is fixed  as  in Eq.\eqref{phifix2},  the squeezing spectra can be written as
\begin{align}
\Sigma_{\pm}^{\text{\tiny{ S}}} (\Om) 
&=\frac {   \sqrt{g^2 + \tilde{\Om}^2}  - g \sin{ \sqrt{g^2 + \tilde{\Om}^2}}   }
{ \sqrt{g^2 + \tilde{\Om}^2}  + g \sin{ \sqrt{g^2 + \tilde{\Om}^2}}  }
\label{analytic}
\end{align} 
where $\tilde \Om= \Omega/\OmGVS$. The antisqueezing spectra, for phases orthogonal to those in Eq.\eqref{phifix2},  are just the inverse
$\Sigma_{\pm}^{A} (\Om) = 1/ { \Sigma_{\pm}^{S} (\Om)}$.
This expressions take a particularly simple form in the neighborhood of threshold and for small frequencies. Let us define a distance from threshold 
$
\epsilon= g_{\mathrm{thr}} -g
$
and let us consider the limit  $\epsilon \ll 1$ and  $|\tilde \Om | \ll g$. By expanding the various functions in Eq.\eqref{analytic}  around $ \epsilon=0$ and $\tilde \Om  / g =0$, and keeping terms at most quadratic in the small quantities, we obtain
\begin{align}
\Sigma_{\pm}^{ S} (\Om) \underset{   \substack{    { \epsilon \ll1} \\ {|\tilde \Om | \ll g}  }     }{\longrightarrow} \; &
\frac{1}{4}    \left(\epsilon^2  +   \frac{  \tilde{\Om}^2 }{g_{\mathrm{thr}}^2}     \right)\, .
\label{epsilon}
\end{align} 
This function  is a parabola which reaches its  minimum at $\Sigma_{\pm} (0) =\frac{\epsilon^2}{4} \to 0 $ as $\epsilon \to 0$,  and of width 
$\Delta\tilde  \Omega \approx 2g_{\mathrm thr}$ constant close to threshold.
In the same limit,  the antisqueezing spectra become
\begin{align}
\Sigma_{\pm}^{ A} (\Om)  \underset{   \substack{    { \epsilon \ll1} \\ {|\tilde \Om | \ll g}  }     }{\longrightarrow}\; &
\frac{4} {    \epsilon^2  +   \frac{  \tilde{\Om}^2 }{g_{\mathrm{thr}}^2}     }
\label{antiepsilon}
\end{align} 
which represents a Lorentzian peak  of diverging height $\frac{4}{\epsilon^2} \to \infty$  and  of vanishing width $ \Delta \tilde \Om = \epsilon g_{\mathrm thr} \to 0$,  as threshold is approached. These approximated formula nicely  reproduce  the minima and the maxima  in Figs.\ref{fig_squeezing} and \ref{fig_antisqueezing}, respectively, when  not too far from  threshold, and are actually valid 
 for rather large distances from threshold, as shown by Fig.\ref{fig_sigma0} . 
\par
We notice that such behaviors of the squeezing spectra  are completely  comparable to what can be obtained in standard cavity OPOs below threshold (see e.g  Ref.\cite{Walls1994}, formula (7.59),page 131).  Here, in the degenerate case,   the spectrum of squeezing has the form 
\begin{align}
\Sigma^{\mathrm OPO} (\Om) &= \frac     {(A_p^{thr} -A_p)^2  + \bar \Omega^2}{(A_p^{thr} +A_p)^2  + \bar \Om^2} 
\underset  {|\bar \Omega |\ll 2} {\to}\frac{1}{4}    \left(\epsilon^2  +   \bar{\Om}^2  \right)
\end{align}
where $A_p$ is a cavity gain parameter, proportional to the pump amplitude, the $\chi^{(2)}$ susceptibility and the photon lifetime in the cavity;  $\bar \Om$ is the frequency normalized to the cavity linewidth;  the OPO threshold is at $A_p= A_p^{thr}=1$, and $\epsilon 
= A_p^{thr} -A_p $ defines also in the OPO case  the dimensionless distance from threshold. Remarkably, in both MOPO and OPO cases, 
the
behavior $\sim \epsilon^2 /4 $  with the distance from  threshold   indicates that excellent level of squeezing can be obtained even at rather large distances below the  threshold. 
\par
As a final,  we remark that the spectra in Figs.\ref{fig_squeezing} and \ref{fig_antisqueezing} were calculated  in the specific  case of  periodically poled Lithium Niobate ( PPLN), pumped at $800$nm,  with $\Lambda= 368$ nm, suitable to phase-match the type 0 process at $\lambda_s= \lambda_i$.   The wave-numbers were evaluated using the complete Sellmeier relations in \cite{Nikogosi͡an2005}. However, we did not notice  appreciable differences (unless at very large frequencies $\Om \ge 15 \OmGVS$) with the linearly approximated results \eqref{analytic}, nor with curves obtained for different materials or  tuning conditions, which confirms  that  our results are completely general for any MOPO configuration. 
\par 
 In conclusions, the MOPO below threshold  is a source of  EPR entangled beams
over a wide range of light frequencies, including telecom wavelengths.  Our analysis has shown that this cavityless configuration of PDC  
can reach the same narrowband, high level, and robust correlation characteristic of the  cavity OPO, which represents the golden standard to for EPR beams. As such, it can be used as an alternative to the OPO, meeting the increasing demand for monolithic devices in the field of  integrated quantum optics. 


\bibliography{biblio}
\end{document}